# COLLABORATIVE LEARNING MODEL WITH VIRTUAL TEAM IN UBIQUITOUS LEARNING ENVIRONMENT USING CREATIVE PROBLEM-SOLVING PROCESS

Sitthichai Laisema[1] and Panita Wannapiroon[2]

[1]Ph.D student, Information and Communication Technology
for Education Division, Faculty of Technical Education,
King Mongkutt's University of Technology North Bangkok, Thailand
[2]Assistant Professor, Information and Communication Technology
for Education Division, Faculty of Technical Education,
King Mongkutt's University of Technology North Bangkok, Thailand

*ABSTRACT*

*The purposes of this research study were: 1) to develop a Collaborative Learning Model with Virtual Team in u-Learning Environment using Creative Problem-solving Process(U-CCPS Model); 2) to evaluate a U-CCPS Model. The research procedures were divided into two phases. The first phase was to develop U-CCPS Model, and the second phase was to evaluate U-CCPS Model. The sample group in this study consisted of five experts using purposive sampling. Data were analyzed by arithmetic mean and standard deviation. The research findings were as follows: The U-CCPS learning Model consisted of five components as follows: 1) Input factors, 2) Process, 3) Control, 4) Output and 5) Feedback. The input factors consisted of four components as followed: 1) Objectives of U-CCPS Model, 2) Roles of Instructors, 3) Roles of learners and 4) Design of learning media. The process consisted of two components as followed: 1) Preparation before learning, and 2) Instructional management process. The experts agree that a U-CCPS Model was highest suitability.*

## 1. INTRODUCTION

Partnership for 21st Century Skills has developed some visions for the success of learners in economic systems of the new world, and for the practitioners to integrate skills with the teaching of academic contents. Accordingly, the 21st Century Student Outcomes and Support Systems were established by the combination of knowledge, specialized skills, expertise, and omniscience, all of which contribute to the success in both career and life[1]. The skills that can insure the learner's preparedness to such a complicated working life as seen in this modern era include Creativity and Innovation, Critical Thinking and Problem Solving, and Communication and Collaboration [1].

As to the dynamic changes of the world in the 21st century, those who can survive and blend themselves with this modern society must practice their creative thinking. Creative thinking already exists in everybody, but it will be more acute, more active and more sustainable with proper learning and practice. Those equipped with high creative thinking skills will always receive better jobs, live more prosperous life, and do anything more helpful to the world [2].

In addition to creative thinking, collaborative skill is also indispensable for the learners of 21$^{st}$ century. This skill refers to the ability to work efficiently with others in a team, while having flexibility in their own roles and dedication towards the team's tasks in order to finally achieve the mutual goals. Thereby, in order to work well with others, the learners must have responsibility in collaboration and recognize other team members' performance.

The instructional technique based on Creative problem-solving process is so popular that most educators from different institutes have applied it to both adult and youth education [3]. This is because this technique can be easily used in daily life, and easy to learn and understand in all age groups, all situations and all cultures. It is also practical when applied to solve common problems, and above all, it is a technique designed especially to develop problem-solving skill and creative thinking skill [4].

Several scholars have studied and developed the creative problem-solving process. Isaksen, Dorval and Trafinger developed the model of creative problem-solving process, in which they made languages and process thereof more flexible so that it could better solve the problems in different contexts [5]. Once having been tested and used by Maraviglia and Kvashny, the new developed creative problem-solving model was found to have the highest influences on the development of creative thinking [6].

According to National ICT Policy Framework 2011-2020: (ICT2020), Strategy 6, the objective thereof is to reduce the economic and social inequality by creating the equal access to public resources and services, particularly the fundamental services necessary for life and for the well-being of citizens [7]. The said strategy will also promote the production and application of all innovative digital learning media, as well as the publication of all electronic media or lessons. Meanwhile, the said policy framework also places an emphasis on the use of information and communication technology in education management.

U-learning is an instructional management in which information and communication technology is used in education management. Since this kind of learning is based on Ubiquitous technology, which can create learning in different environments as to the contexts of learners [8], the learners are able to learn anywhere anytime they want through their mobile device with no need of access to the computer. Consequently, there are flexibility in learning and quick access to desired information. It is also a kind of learner-centered instructional management which focuses on the works of learners. So, this kind of learning will encourage learners to create and acquire knowledge by themselves [9].

U-learning can be applied with Constructivism [10]. The use of learning theory to design education can well link knowledge information of learners to environments (Jacobs, 1999). The researcher is interested in collaborative learning, which enables learners of different ability to cooperate in teamwork. The team members will study something of common interest by creating a project. Then, the team will present the knowledge they have got from the said collaborative learning. Also, during the process of managing knowledge information, creating a project together, the learners have to exchange their opinions. This is the highlight of collaborative working, which depends mostly on collective action and mutual understanding [11]. Furthermore, collaborative learning helps learners do their job with high efficiency by combining parts of their existing knowledge and synthesizing them to create the new knowledge [12],[13]. Therefore, collaborative learning is an instructional management that promote the learners to have collaboration skills [14].

Collaborative learning encourages the learners to work in a team in order to create a work piece as required by the instructors. Thereby, this collaborative learning enables the learners to create a virtual team so that the members can contact one another through the media with no face-to-face meeting. Although the members do not stay in the same place [15],[16],[17], the virtual team can still help them work together all 24 hours, providing the team with more various skills and different perspectives. To summarize, collaborative learning in the virtual team, in spite of different places and time, helps the learners work or learn together by means of information and communication technology [17],[18].

It is therefore necessary to develop a learning model with the application of information technology in order that the learners could have both creative thinking skill and collaboration skill. Thus, the researcher is interested in the study of collaborative learning with virtual team via electronic media based on Creative problem-solving process in Ubiquitous learning environment so that the learners could develop creative thinking skill and collaboration skills.

## 2. PURPOSE OF THE STUDY

The purposes of this study were;

**2.1.** To design a develop a Collaborative Learning Model with Virtual Team in Ubiquitous Learning Environment using Creative Problem-solving Process to enhance Creative Thinking and Collaboration Skills.

**2.2.** To evaluate a Collaborative Learning Model with Virtual Team in Ubiquitous Learning Environment using Creative Problem-solving Process to enhance Creative Thinking and Collaboration Skills.

## 3. SCOPE OF THE STUDY
### 3.1. Population

The study population was experts in instructional design, ubiquitous Learning, information technology, creative thinking and collaboration skills.

### 3.2. Sample groups

The sample groups of study were five experts in instructional design, ubiquitous Learning, information technology, creative thinking and collaboration skills by purposive sampling.

### 3.3. Variables of the study

An independent variable was Collaborative Learning Model with Virtual Team in u-Learning Environment using Creative Problem-solving Process. A dependent variable was evaluation of the proposed learning model.

## 4. CONCEPTUAL FRAMEWORK

The conceptual framework of this study is shown in Figure 1. five components were used for creating a U-CCPS model which theoretically affected the Collaborative Learning with Virtual team, Creative Problem-Solving Process, Ubiquitous Learning environment, Creative thinking and Collaboration Skills of the learner.

### 4.1. Collaborative learning with virtual team

Collaborative learning with virtual team refers to the method that enables the learners to cooperate in the team working in order to study something of common interest by creating a project and then presenting the knowledge they have got from the said collaborative learning. The learners in the team study and create knowledge together; thereby, with the collaborative learning with virtual team, they can still work or study together, despite different places and time, by the aid of information and communication technology [13],[15],[16],[17],[18],[19].

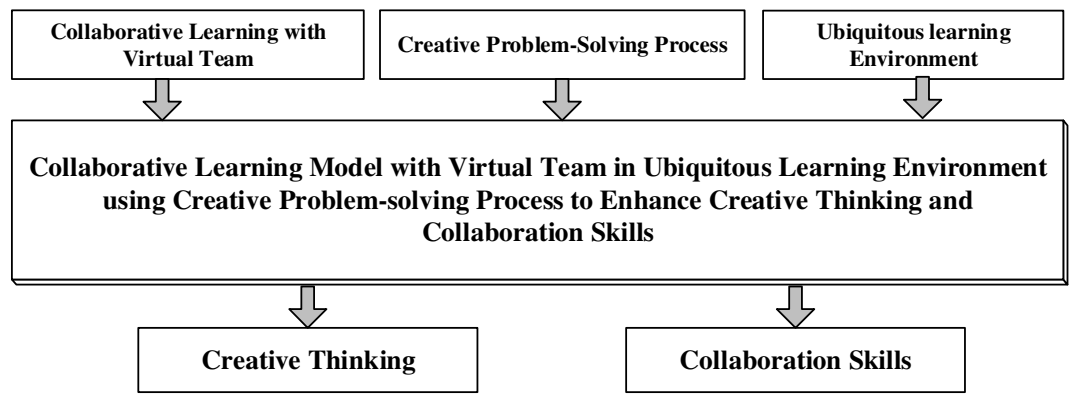

Figure 1. conceptual framework of U-CCPS Model

## 4.2. Creative problem-solving process

Creative Problem-Solving Process is a method to combine creative thinking with experiences and information research in order to find out the solutions. There are 4 main steps in solving the problems [3],[4],[5].

### 4.2.1. Understanding the problem

In creative problem-solving process, once we completely understand the problem or really know its context, it is easy to find out the solution thereof.

### 4.2.2. Generating ideas

To generate any ideas to find out the solution or answer to the questions of the previous step, the extraordinarily new and different ideas are needed.

### 4.2.3. Planning for action

T In this step, there are solution finding and acceptance finding. The first one is about analysis, definition, and adaptation of the ideas to be more concrete based on elaborate consideration and examination. The other is about the finding of support and objection in order to bring about the solution.

### 4.2.4. Appraising tasks

Creative problem solving process is effective and flexible, and it can be adjusted to suit any individuals, problems or situations.

## 4.3. Ubiquitous learning environment (ULE)

Ubiquitous Learning is a kind of learning in a form of digital media, in which the learners can study anything anytime and anywhere without a computer. As a result, there are flexibility in learning and fast access to the information; whereby the learning will be in accordance with different environments and contexts of the learners. The characteristics of ULE include [10],[20],[21],[22],[23],[24],[25],[26]:

### 4.3.1. Computer tablets

Computer tablets with processing unit and memory are equipped with a system that can check the status of learners before sending them the contents through the said tablets.

### 4.3.2. Wireless Communication

Wireless Communication, e.g. Bluetooth, 3G or Wifi, is suitable for the fast data transfer. This research employs Wifi network or 3G, in which the learners can study anything wherever there is Wifi or 3G available.

### 4.3.3. Ubiquitous Learning Mangement System (U-LMS)

U-LMS has a host computer for the management of learning, and storage of education resources, media, and education units. The host computer can also provide the learners with understanding and assistance by analyzing and answering the learners' questions through their tablets.

### 4.3.4. Context Awareness

Context Awareness will detect the movement and environment according to the learners' context so as to recognize their status.

## 4.4. Creative Thinking

Creative Thinking refers to the advanced cognitive process that employs several thinking processes to create new things or solve existing problems. Creative thinking exists only when the creators have freedom of thinking, or Divergence Thinking, and ability to adapt or combine the existing thinking, which will lead to the discovery of novelties. Divergence Thinking is a kind of creative thinking. It means the ability of an individual to solve the problems, the ways of thinking that generate different and new things, and the ability that can be applied in different kinds of jobs. Divergence Thinking consists of [27]:

### 4.4.1. Fluency

Ability to think and respond to stimuli at the best, or to find out the right but different solutions in the same issue.

### 4.4.2. Flexibility

Ability to adjust the thinking and make it comply with different circumstances, focusing on the wide ranges of fluency by means of classification and criteria.

### 4.4.3. Originality

Ability to think differently and uniquely, or to modify existing knowledge so as to create new things.

### 4.4.4. Elaboration

Ability to see through the details that are invisible to others, and to link different things together in a meaningful manner.

### 4.5. Collaboration skills

Collaboration skills consists of [1]:

**4.5.1.** Demonstrate ability to work effectively and respectfully with diverse teams.

**4.5.2.** Exercise flexibility and willingness to be helpful in making necessary compromises to accomplish a common goal.

**4.5.3.** Assume shared responsibility for collaborative work, and value the individual contributions made by each team member.

## 5. METHODOLOGY

### 5.1. The first phase

Collaborative Learning Model with Virtual Team in Ubiquitous Learning Environment using Creative Problem-solving Process to enhance Creative Thinking and Collaboration Skills with the following method:

**5.1.1** To study, analyze and synthesize documents and former research relevant to the elements of Ubiquitous learning, Collaborative Learning with virtual team and Creative Problem-Solving Process. Then, the results thereof were used to set up a conceptual framework in order to develop a model of Collaborative Learning with virtual team.

**5.1.2** To study information about learning management by interviewing the instructors in order to synthesize the data of learning model and by interviewing the students about their ability to use information technology and communication for learning, their learning style, and their cognitive style.

**5.1.3** The development of the model of U-CCPS model in this phase was derived by analyzing the principles of Collaborative Learning with virtual team and Creative Problem-Solving Process. Then, the results of the study were used to identify U-CCPS model based on the following components: 1) Input factors, 2) U-CCPS process, 3) Control, 4) Output and 5) Feedback.

**5.1.4** To present the U-CCPS model to the advisors for consideration and revision.

**5.1.5** To present the U-CCPS model model to the experts for consideration by means of in-depth interview.

**5.1.6** To create the tools for evaluating the suitability of the model of U-CCPS model.

### 5.2. The second phase

The second phase of the project was an evaluation of Collaborative Learning Model with Virtual Team in Ubiquitous Learning Environment using Creative Problem-solving Process to enhance Creative Thinking and Collaboration Skills, with a method as follows:

**5.2.1** To present the developed activity to the five experts from the fields of instructional design, ubiquitous Learning, information technology, creative thinking and collaboration skills, for suitability evaluation.

**5.2.2**   To improve the model of Collaborative Learning model according to the suggestions of the experts.

**5.2.3** To present the model of Collaborative Learning model in the form of a diagram with report.

**5.2.4** To analyze the results of evaluation of the model by mean ($\bar{x}$) and standard deviation (S.D.) consisting of five criteria for evaluation, according to the ideas of Likert; that is highest, hight, moderate, low and lowest.

## 6. COLLABORATIVE LEARNING MODEL WITH VIRTUAL TEAM IN UBIQUITOUS LEARNING ENVIRONMENT USING CREATIVE PROBLEM-SOLVING PROCESS TO ENHANCE CREATIVE THINKING AND COLLABORATION SKILLS (U-CCPS MODEL)

U-CCPS Model was developed on the basis of System Approach, consisting of 5 main elements and 10 sub-elements.

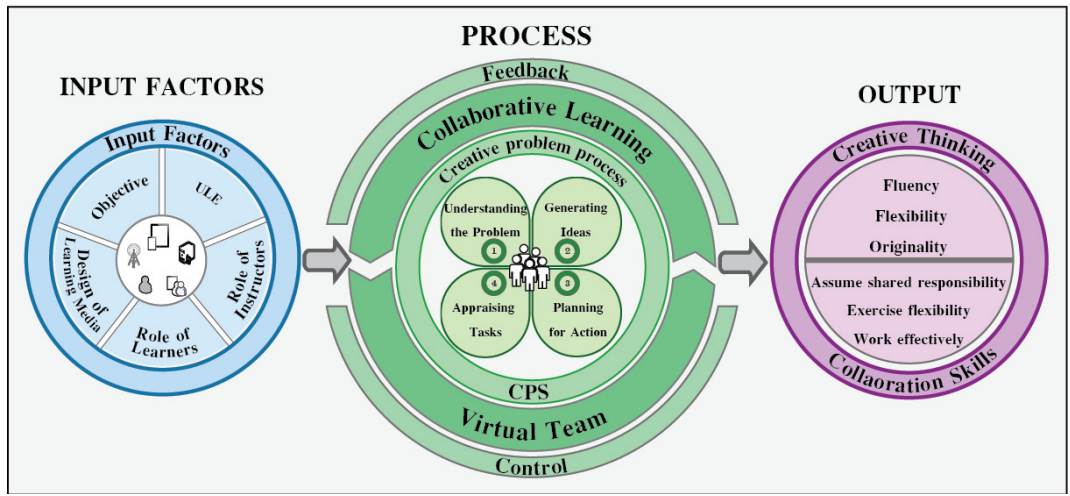

Figure 2.  Collaborative Learning Model with Virtual Team in Ubiquitous Learning Environment using Creative Problem-solving Process to enhance Creative Thinking and Collaboration Skills (U-CCPS Model)

### 6.1. Input Factors

The input factors consisted of four components as followed: 1) Objectives of U-CCPS Model, 2) Roles of Instructors, 3) Roles of learners and 4) Design of learning media.

### 6.1.1. Objectives of U-CCPS Model

The objectives of this model are to develop creative thinking in terms of fluency, flexibility, elaboration, and collaboration skill.

### 6.1.2. Ubiquitous Learning Environment(ULE)

Ubiquitous learning is a seamless learning whenever it is in information space or in physics space, through ubiquitous computing information space and physics space are converged. In ULE Learning, learning demands and learning resources are everywhere; study, life and work are connected each other. When learners meet any practice problem ubiquitous computing help them to resolve it at anytime, anywhere. The learners can easily perception and obtaining learning objects detailed information and content through situational perception of mobile devices. Using dialogue, living community, cooperation studies, social process of internalization, participate in joint activity to realize social learning. Happen of effective ubiquitous learning depends on founding of learning environment [28].

### 6.1.3. Roles of Instructors

The instructors control the learning, maintain it in the learning process, provide advice and assistance, and design learning activities in order to enhance learners' creative thinking and collaboration skills. The instructors also have to prepare instructional media, contents, and management system thereof. Moreover, the instructors have an important role in Ubiquitous learning environment, e.g. speaking with learners, holding seminar, questioning and answering problems with learners, following up the learners, giving feedback information to the learners, and checking the works of learners.

### 6.1.4. Roles of learners

The learners do learning activities that are designed by the instructors. The said learning activities encourage the learners to work in groups; whereby the learners have to work with the other members. However, each of them does not have to meet because they can work through their mobile devices. In the learning process, the learners are asked to produce a piece of work together as to the topic provided by the instructors

### 6.1.5. Design of learning media

The learning media herein are in the form of online learning media that can be modified, changed, and publicized very quickly. It can also be accessed with no limitation of duration, places, or devices. The learners can choose any contents to learn or any activity to do based on their own interest. The learning media can be displayed on any kind of mobile device.

## 6.2. Process

The process consisted of two components as followed: 1) Preparation before learning, and 2) Instructional management process.

### 6.2.1. Preparation before learning
#### 6.2.1.1 Orientation

This activity was held to provide knowledge and understanding of learning activities, evaluation, tests, and collaborative learning with virtual team in u-learning environment using creative problem-solving process.

#### 6.2.1.2 Register

Register through Ubiquitous Learning Management System - All learners had to register in U-LMS to participate in learning activity of the system.

#### 6.2.1.3 Group of the learner

The learners were divided into groups based on their interests, and they could use any tools to communicate, to work together, and to analyze any problems.

### 6.2.1.4 Test of creative thinking and collaboration skills

Test of creative thinking and collaboration skills to measurement of the scores before taking part in the developed learning model.

### 6.2.2. U-CCPS process

The researcher developed U-CCPS Model with an instructional management process, in which creative thinking was combined with experiences and researches in order to figure out different solutions. The instructional management process using Ubiquitous learning environment includes 4 main steps and 7 sub-steps.

### 6.2.2.1 Understanding the problem

In CPS Process, the true understanding of problems or the contexts thereof will facilitate the finding out of solutions. In this process, there are 3 sub-steps as follow:

**Constructing opportunities**

In this step, the learners will find out the problems by themselves. The group members mutually set up general target that leads to problem solving. The instructors provide the topic for the learners to search all relevant problems. The members of the group help one another to search for the problems on the basis of their experiences, roles, and status. Thereby, everybody sees problem solving as a way to construct opportunities. U-LMS system proposes a topic or an issue in order for the learners to find out problems. Then the learners of each group try to search for the problems as to their own knowledge and experiences. The tool used here is Group Discussion via web board and online chat. The learners can work together even though they are in different places and different contexts. When the group members post an issue on the web board, the other members who have similar opinions will leave comments in the said issue so that there would be as many issues as possible from the members. If the other members have different issues to post, they can do so immediately. Whenever the group members leave their comments in any issues, the system will notify the other members about this automatically. In this way, the learners can still work together through their mobile device although they are in different places and different contexts.

**Exploring data**

This step focuses on the collection of facts, opinions, satisfaction, confirmation, conflicts, and consideration of overall contexts. The questions are to be posted to get the clarity of issue contexts. Information, understandings, and feelings are also collected in this step to better understand the issues. In other words, those who want to solve the problems have to search for information and classify it (convergence) so as to obtain the information most effective to the problem solving. Each member of the team, in this step, is collecting data relevant to the issues posted on the web board and online chat of U-LMS system. This is to finalize the most important information that can define the said issues.

**Framing problem**

This step is designed to specially help the problem solvers and to make the issues more obvious. Thereby, the group members consider and choose the most significant issues, based on the concept provided by instructors, out of all that have been posted. In this step, the members have to cooperate via the web board and group discussion. The tool used here is Google Document on U-LMS System.

#### 6.2.2.2 Generating ideas

This step is to find out the ways to solve the problems or answer the questions in the previous steps. The questions in this step have no fixed answers, so there must be divergent thinking in terms of fluency, flexibility, and elaboration. Then, the ideas are collected and the one with highest possibility will be chosen. That is why this step is called "generating ideas". The highlight in this step is the opportunity to extend thinking, both in and out of the box. The learners help one another to look for as many solutions as possible by posting them on their group web board. Besides, the learners search for information, brain storm, and share the data with the other members by means of Google Document on their mobile devices. This is to create online documents derived from their collaboration.

#### 6.2.2.3 Planning for action

This step includes:

**Solution finding**  This step is to analyze, define, and clarify the ideas so that they are more concrete. It also requires meticulous consideration and investigation. The main point of this step is to scrutinize and select the likely practical option that is manageable and most efficient to the tasks. The highlight of this step is that all group members choose the tools or methods that are really practical by picking out the best idea in order to attain the best solution. The tools here include the group chat where the members can share their ideas, and the web board to post their opinions and choose the optimal solution.

**Acceptance finding**  This step is to find acceptance and opposition so as to derive solutions. The selection of solutions is based on individuals, places, equipment, or time that can promote the operation plan to be successful. The management must be proceed as planned; e.g. the members must strictly follow the plan, attend the meeting, and provide acceptance for the solutions. In this step, the group members work together to set up an operation plan and find out the solutions with the aid of Google Document on their mobile devices.

#### 6.2.2.4 Appraising tasks

Creative problem-solving process is so effective and flexible that it is chosen as a tool to find out solutions. After the members have produced a work piece as planned, their tasks and their work will be appraised by instructors, friends, and experts. Whereby, the said appraisal is done through U-LMS System on mobile device.

### 6.3. Control

Learning activities based on U-CCPS Model encourages the learners to have creative thinking and collaboration skills, and to search for knowledge by themselves via U-LMS System provided by the instructors. The learners have to work together, solve the problems, and share their knowledge under the control of the system that is provided by instructors. The control is divided into:

#### 6.3.1. Control and Follow up of learners
U-LMS will examine and follow the learners in every aspect such as interest in study and activity, enthusiasm in problem solving and finding solutions, determination and responsibility for work. The system also provides feedback information.

#### 6.3.2. Set up timeline to examine the tasks

U-LMS will set up a timeline to check out the tasks and use it as a tool to control and encourage the learners to study and do all activities in due time.

International Journal on Integrating Technology in Education (IJITE) Vol.2, No.4, December 2013### 6.4. Output

Measurement and assessment in each unit are all authentic. The learning outcomes, Torrance creative thinking, and collaboration skills will be measured after collaborative learning with virtual team activities using creative problem-solving process in ubiquitous learning environment.

### 6.5. Feedback

This refers to the analysis of information from different steps of U-CCPS Model, opinions of the experts, and opinions of the learner. This is to optimize all steps of the instructional model and to achieve the goals as expected.

## 7. THE EVALUATION RESULTS OF THE U-CCPS MODEL

The evaluation is carried out by submitting the developed model to the five experts for a certification on the suitability of its components, U-CCPS process, and for a test. The evaluation result by the expert has shows that the components have highest suitability ($\bar{x}$ = 4.52, S.D. = 0.58), the U-CCPS process have the highest suitability ($\bar{x}$ = 4.58, S.D. = 0.65), and the overall appropriateness for a test have the high suitability ($\bar{x}$ = 4.47, S.D. = 0.55).

Table 1. The evaluation results of U-CCPS Model.

| Evaluation Lists | Results | | Level of suitability |
|---|---|---|---|
| | $\bar{x}$ | S.D. | |
| 1. Input factors | 4.80 | 0.45 | Highest |
| 2. Process | 4.60 | 0.55 | Highest |
| 3. Control | 4.40 | 0.55 | High |
| 4. Output | 4.20 | 0.45 | High |
| 5. Feedback | 4.60 | 0.89 | Highest |
| **Summary** | **4.52** | **0.58** | **Highest** |

The table 1. Shows that the experts agree that a U-CCPS components was highest suitability. ($\bar{x}$ = 4.52, S.D. = 0.58)

Table 2. The evaluation results of Instructional Process

| Evaluation Lists | Results | | Level of suitability |
|---|---|---|---|
| | $\bar{x}$ | S.D. | |
| **1. Preparation before learning** | **4.70** | **0.59** | **Highest** |
| 1.1 Orientation | 4.60 | 0.55 | Highest |
| 1.2 Register | 4.80 | 0.45 | Highest |
| 1.3 Group of learners | 4.80 | 0.45 | Highest |
| 1.4 Test of creative thinking and collaboration skills | 4.60 | 0.89 | Highest |
| **2. U-CCPS Process** | **4.45** | **0.71** | **High** |
| 2.1 Understanding the problem | 4.60 | 0.89 | Highest |
| 2.2 Generating ideas | 4.20 | 0.84 | High |
| 2.3 Planning for action | 4.60 | 0.55 | Highest |
| 2.4 Appraising tasks | 4.40 | 0.55 | High |
| **Summary** | **4.58** | **0.65** | **Highest** |

The table 2. shows that the experts agree that a U-CCPS Process was highest suitability. ($\bar{x}$ = 4.58, S.D. = 0.65)

Table 3.  The evaluation results of U-CCPS model for a test

| Evaluation Lists | Results | | Level of suitability |
|---|---|---|---|
| | $\bar{x}$ | S.D. | |
| 1. U-CCPS Model is appropriate to develop creative thinking | 4.60 | 0.55 | Highest |
| 2. U-CCPS Model is appropriate to develop collaboration skills | 4.40 | 0.55 | High |
| 3. U-CCPS Model is possible for test | 4.40 | 0.55 | High |
| **Summary** | **4.47** | **0.55** | **High** |

The table 3. shows that the experts agree that a U-CCPS Model was appropriateness for test in the high ($\bar{x}$ = 4.47, S.D. = 0.55).

## 8. DISCUSSION

**8.1.** The results of assessment on the model's elements show that the 5 main elements of the model; i.e. Input factors, Process, Output, Control and Feedback, are highest suitable. This is because the process of instructional design employed the principle of analysis, design, development, implement and evaluate (ADDIE Model). The results are in accordance with the research of Tekinarslan et al., [29] who found that the instructional design of ADDIE could be applied in Ubiquitous instructional design. In addition, this research used the concepts and principles of collaborative learning in Ubiquitous environment. This is in compliance with the research of Tseng et al.,[30] who found that collaborative learning in Ubiquitous environment encouraged cooperation of learning among the learners everywhere and every time, and it also complied with the learning contexts of learners.

**8.2.** The results of assessment on the instructional model at the preparation step before learning show that the model is highest suitable. This is because Ubiquitous instructional design is quite new and it requires orientation to get the learners well prepared for learning. The results are in line with the research of Olaham et al.,[31] who found that there should be orientation before any learning so that the learners could understand different aspects and be able to study in Ubiquitous environment as efficiently as possible..

**8.3.** The results of assessment on the instructional process at the management step show that the process is highest suitable. The preparation step included the steps of understanding the problem, generating ideas, planning for action, appraising tasks, and summarizing the principles and concepts. This corresponds to the research of Treffinger, Isaksen and Dorval [5] , who found that the instructional activities using creative problem-solving process had to include the steps of understanding the problem, generating ideas, planning for action, and appraising tasks.

**8.4.** The results of assessment on the application of instructional model by the experts show that the model is highest suitable. This is because the instructional model as well as its steps and activities of collaborative learning were all suitable to the creative problem-solving skill. It complies to the research of Maraviglia and Kvashny [6], who applied the instructional model using creative problem-solving process. They found that the newly developed creative problem-solving process had highest influence on the deveopment of creative thinking and creative problem-solving skills. This also corresponds to the research of Suangsuda Pansakul [11], who studied and presented the learning model of creative and collaborative problem-solving process in the internet-based organizations. She found that the sample group, after participating in the learning process, had higher level of creative problem-solving skills than they did before. Moreover, the research of Lau et al. [14] found that collaborative learning is the process that provides the learners with collaboration skills.

**Authors**

**Sitthichai Laisema** received a Bachelor of Education(Hons) from Silpakorn University in 2006; Master of Sciene in Industrial Education in the field of Computer and Information Technology from KMUTT, Thailand in 2008. Currently, he is a Ph.D. candidate in Information and Communication Technology for Education at Division of Information and Communication Technology for Education, Faculty of Technical Education, King Mongkut's University of Technology North Bangkok (KMUTNB) and lecturer at Department of Educational Technology, Faculty of Education, Silpakorn University, Thailand. He has research experience in information and communication technology for education.

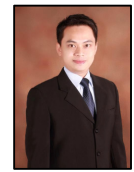

**Dr.Panita Wannapiroon** is an Assistant Professor at Division of Information and Communication Technology for Education, Faculty of Technical Education, King Mongkut's University of Technology North Bangkok (KMUTNB),Thailand. She has experience in many positions such as the Director at Innovation and Technology Management Research Center, Assistant Director of Online Learning Research Center, Assistant Director of Vocational Education Technology Research Center, and Assistant Director of Information and Communication Technology in Education Research Center. She received Burapha University Thesis Award 2002. She is a Membership of Professional Societies in ALCoB (APEC LEARNING COMMUNITY BUILDERS) THAILAND, and Association for Educational Technology of Thailand (AETT)

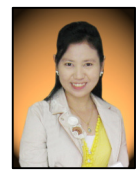